\DocumentMetadata{}
\PassOptionsToPackage{noend}{algpseudocode}
\documentclass[manuscript,anonymous=false]{acmart}
\usepackage[utf8]{inputenc}
\usepackage{enumitem}
\usepackage{subfig}
\usepackage{url}
\usepackage{multirow}
\usepackage{graphicx}
\usepackage{booktabs}
\usepackage{tabularx}
\newcolumntype{b}{X}
\newcolumntype{s}{>{\hsize=.25\hsize}X}

\usepackage{xcolor}
\colorlet{punct}{red!60!black}
\definecolor{background}{HTML}{EEEEEE}
\definecolor{delim}{RGB}{20,105,176}
\colorlet{numb}{magenta!60!black}

\usepackage{researchpack}
\algrenewcommand{\algorithmicrequire}{\textbf{Precondition:}}
\algrenewcommand{\algorithmicensure}{\textbf{Output:}}

\definecolor{BLUE}{rgb}{0,0,1}
\def\revcolor{black}
\newcommand{\rev}[1]{\textcolor{\revcolor}{#1}}

\def\revcolorTwo{black} 
\newcommand{\revTwo}[1]{\textcolor{\revcolorTwo}{#1}}

\begin{document}

\title[The AI-Therapist Duo]{The AI-Therapist Duo: Exploring the Potential of Human-AI Collaboration in Personalized Art Therapy for PICS Intervention}

\author{Bereket A. Yilma}
\authornote{Both authors contributed equally to this research.}
\email{bereket.yilma@uni.lu}
\affiliation{%
  \institution{University of Luxembourg}
  \country{Luxembourg}
}

\author{Chan Mi Kim}
\authornotemark[1]
\email{c.m.kim@utwente.nl}
\affiliation{%
  \institution{University of Twente}
  \country{The Netherlands}
}
\author{Geke Ludden}
\email{g.d.s.ludden@utwente.nl}
\affiliation{%
  \institution{University of Twente}
  \country{The Netherlands}
}
\author{Thomas van Rompay}
\email{t.j.l.vanrompay@utwente.nl}
\affiliation{%
  \institution{University of Twente}
  \country{The Netherlands}
}

\author{Luis A. Leiva}
\email{luis.leiva@uni.lu}
\affiliation{%
  \institution{University of Luxembourg}
  \country{Luxembourg}
}

\renewcommand{\shortauthors}{Yilma et al.}

\begin{abstract} 
Post-intensive care syndrome (PICS) is a multifaceted condition that arises from prolonged stays in an intensive care unit (ICU). While preventing PICS among ICU patients is becoming increasingly important, interventions remain limited. Building on evidence supporting the effectiveness of art exposure in addressing the psychological aspects of PICS, we propose a novel art therapy solution through a collaborative Human-AI approach that enhances personalized therapeutic interventions using state-of-the-art Visual Art Recommendation Systems. We developed two Human-in-the-Loop (HITL) personalization methods and assessed their impact through a large-scale user study (N=150). Our findings demonstrate that this Human-AI collaboration not only enhances the personalization and effectiveness of art therapy but also supports therapists by streamlining their workload. While our study centres on PICS intervention, the results suggest that human-AI collaborative Art therapy could potentially benefit other areas where emotional support is critical, such as cases of anxiety and depression. 
\end{abstract}

\begin{teaserfigure}
\includegraphics[width=\textwidth]{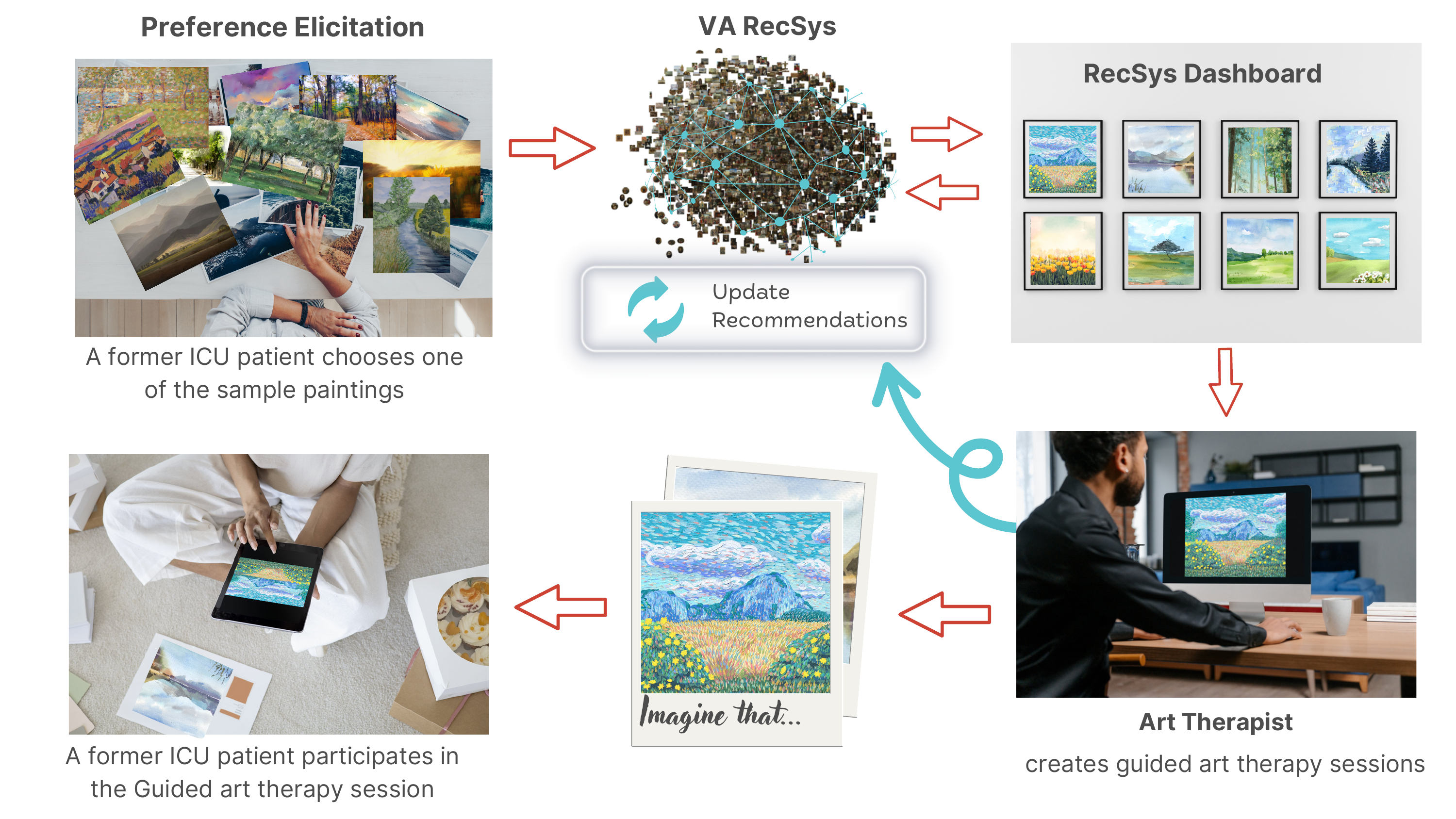}
\caption{
    Overview of our art therapy approach for PICS prevention and treatment.
}
\Description{
    Teaser image depicting the proposed system pipeline. From left to right and top to bottom: Preference elicitation of art paintings, system recommendations, a dashboard that allows the therapist to select the most suitable paintings, and create a guided art therapy session for patients.
}
\label{fig:header}
\end{teaserfigure}



\maketitle


\section{Introduction}

Post-intensive care syndrome (PICS) is a debilitating condition that affects more than 75\% of patients who have survived prolonged stays in intensive care units (ICUs)~\cite{needham2013physical, rawal2017post,pandharipande2013long,davydow2013hospital}. The psychological and cognitive impairments associated with PICS can have a profound impact on patients' quality of life, leading to increased anxiety, depression, and difficulty reintegrating into daily life~\cite{dowdy2005quality,griffiths2013exploration}.

Currently practiced PICS interventions, such as individual counseling or reflections~\cite{cuthbertson2009practical, inoue2019post}, are effective yet can be time-consuming and may not adequately address the unique needs of each patient. 
In recent years, there has been growing interest in Human-Computer Interaction (HCI) to explore innovative methods to prevent and reduce PICS. The examples include a dynamic lighting system~\cite{O'Reilly} and a smart artificial window~\cite{kim2024outside} featuring various nature scenes and creating an adaptive atmosphere to support patients' circadian rhythms and restoration in the ICU. Another promising area is art therapy, which has been shown to significantly improve patients' mood and affective states~\cite{schouten2015effectiveness, yilma2024artful}. Specifically, art therapy that involves nature content hold promise, as nature exposure has been shown to significantly improve the mood and affective states of ICU patients~\cite{ulrich1984view}. While art therapy can take from 8 to 15 weeks to yield significant results, its integration into formal rehabilitation programs remains limited due to several factors~\cite{regev2018effectiveness}. One of these factors is the limited availability of registered art therapists, with only about 7,000 in the United States\footnote{\url{https://www.bls.gov/careeroutlook/2015/youre-a-what/art-therapist.htm}} and approximately 5,000 in Europe.\footnote{\url{https://www.hcpc-uk.org/globalassets/resources/factsheets/hcpc-diversity-data-2021-factsheet--arts-therapists.pdf}} Next to this, the uniqueness of each patient's traumatic experiences necessitates highly personalized interventions, posing a challenge for art therapists. They must select effective artworks from a large number of available pieces and continuously supply new content tailored to each patient’s evolving therapeutic needs.

A recent study~\cite{yilma2024artful} explored the potential of using Artificial Intelligence (AI) enabled Visual Art Recommendation Systems (VA RecSys) to address these challenges. The study showed that AI-generated recommendations not only matched but in some cases even surpassed the quality of those curated by experts , demonstrating the potential of advanced algorithms to enhance therapeutic personalization. However, despite these promising results, the inherent risks of deploying AI-driven RecSys independently in a therapeutic context remain consequential. Without human oversight, such systems may inadvertently produce recommendations that are inappropriate or even harmful, highlighting the critical need for a collaborative approach that integrates expert human judgment with AI capabilities. However, to the best of our knowledge, no prior works have explored a human-AI collaborative approach in art therapy. Therefore, in this work, we set out to answer the critical research question: \textbf{Can a collaborative Human-AI approach in art therapy truly enhance personalization and therapeutic outcomes for PICS patients? If so, What are the potential benefits and challenges of its implementation?}

To investigate this question, we introduce \textit{the AI-Therapist Duo}, a novel human-in-the-loop (HITL) approach that integrates advanced AI-driven techniques in VA RecSys with the expertise of human therapists. This approach is designed to explore the potential of enhancing the therapeutic impact of art therapy while also empowering therapists by allowing them to select the best paintings from manageable options, thereby reducing their workload. \rev{ The objective is to create guided therapy sessions that incorporate techniques from narrative therapy, as discussed in Madigan's work on narrative approaches in therapy~\cite{madigan2011narrative}. This involves a multi-phase process that begins with preference elicitation, where patients identify a painting that resonates with their healing journey. Followed by finding a set of paintings that align with appropriate healing concepts to support the patient’s recovery. Here we employ AI, particularly VA RecSys algorithms to provide personalized recommendations. The final step is curating a guided therapy session tailored to the patient’s needs.} By combining human insight with AI capabilities, our approach seeks to enable more personalized and effective therapeutic intervention and to assess its impact on patients, highlighting the added value for therapists.

In sum, this paper makes the following contributions:
\begin{itemize}
    \item We develop and study two HITL personalization methods for VA Recsys in art therapy,
    one visual-only (using ResNet-50 as backbone architecture) and another multimodal (using BLIP).
    
    \item We conduct an expert evaluation to assess the appropriateness of VA RecSys engines and a large-scale study with 150 post-ICU patients to assess the efficacy of the proposed HITL VA RecSys engines. 

    \item We contextualise our findings and discuss the potential of a Human-AI collaborative approach to enhance personalization in art therapy and streamline therapists' workflows for PICS intervention as well as possible applications in other areas requiring emotional support.
\end{itemize}

\section{Related work}
\label{sec:related-work}

\subsection{Art therapy and its role in mental health}
\label{subsec:ATMH}

\rev{
Art therapy has been recognized as an effective tool for promoting mental health in the clinical context~\cite{schouten2015effectiveness}. 
Common forms of art therapy include creative activities such as drawing, painting, and sculpting. In these activities, artwork functions as a non-verbal medium for individuals to express emotions, process trauma, and engage in self-reflection, helping them to manage a range of psychological issues, including anxiety, depression, and Post-traumatic stress disorder (PTSD)~\cite{schouten2015effectiveness,haeyen2020benefits}. Such an approach, however, requires the engagement of expert \rev{therapists} throughout the session and preparation, as well as specific conditions for patients to participate in creative activities. This restricts access for individuals with physical or contextual limitations.
}

\rev{
Visual engagement with artwork is another widely used form of art therapy in clinical settings, particularly for its effectiveness as a positive distraction~\cite{nanda2008undertaking, ulrich2003healingarts}. It offers a broad range of applications for patients due to its less complex and demanding nature. For instance, visual exposure to art has been used for critically ill patients to address stress, anxiety, and perceived pain~\cite{ulrich2003healingarts}, or to alleviate agitation and anxiety in patients with severe mental health conditions~\cite{nanda2011effect}. It has also been applied in operating rooms and recovery areas, where it not only reduced anxiety but also contributed to faster recovery times~\cite{diette2003distraction}.
}
\rev{
The effects of visual engagement with artwork are further exemplified in recent studies~\cite{kim2025towardshealing, yilma2024artful}, which employed a technique from narrative therapy~\cite{madigan2011narrative} to encourage prolonged and active engagement with the artwork. These studies demonstrated significant temporary mood regulation effects for former patients, including those who had been in the ICU. This shows promising potential for visual art engagement as an additional intervention for preventing PICS, alongside currently practiced methods such as ICU diary~\cite{inoue2019post} and post-ICU clinic~\cite{ramnarain2021post, inoue2019post}.
}

\rev{
In art therapy through visual engagement, it is crucial to use appropriate content in the artwork, much of which is dominated by nature-based themes, such as landscapes featuring trees and water~\cite{ulrich1993effects}. The therapeutic effects of nature-based visual art are rooted in theories such as evolutionary theory~\cite{appleton1975experience} and the Biophilia hypothesis~\cite{wilson1986biophilia}. These theories suggest that natural environments, which provided strategic survival advantages for early humans, elicit positive affective responses (for an overview of nature-based visual stimuli eliciting positive affective responses, see Kim et al.~\cite{kim2023morning}). Following the positive outcomes of art therapy through visual exposure there is growing interest in exploring how art experiences might aid in the emotional and psychological recovery of PICS patients~\cite{luo2024emergence}. However, the content of the artwork can also have negative effects if not carefully chosen. A study by Ulrich~\cite{ulrich1993effects} demonstrated that abstract art with straight-edged forms can provoke strong negative reactions explaining that in highly stressful situations, patients may project their negative emotions onto such artwork, leading to adverse visual experiences. These examples underscore that while art therapy holds promises, its application requires careful customization to ensure both its effectiveness and safety. It highlights the importance of thoughtful selection and integration of artwork in a therapeutic practice, which can make its preparation and implementation time- and labour-intensive for the therapist.
}

\rev{\subsection{State-of-the-Art Digital Art Therapy}}
\rev{
Recent advancements in digital technology can offer significant benefits in addressing the limitations of art therapy. 
On the one hand, for example, digital technology supports therapists with efficient scheduling and management of sessions 
while supporting patients to overcome their physical and contextual barriers,
by enabling patients to receive therapy in the comfort of their own homes.
On the other hand, digital technology also allows therapists to address anxieties or compulsions as they arise, 
thereby providing immediate support and guidance~\cite{robledo2021therapy}. 
Digital technology also expands the tools available for art therapy: 
the use of digital tools for creating and appreciating art, 
as well as the introduction of innovative methods such as interventions based on videos or digital games~\cite{shojaei2024exploring}. 
By adopting and integrating these technologies, art therapy can empower patients in both physical and mental manners, 
enabling new avenues for self-expression and therapeutic engagement~\cite{shojaei2024exploring}.
}

\rev{
A recent study by Cooney and Menezes~\cite{cooney2018design} developed a robotic therapist 
that assists individuals in expressing emotions through stylized and symbolic means facilitating self-exploration, 
which might otherwise be limited by personal skill,
while allowing for a highly personalized therapeutic experience.
}\rev{Another study by Liu, Zhou, and An \cite{liu2024he} integrated the use of image-based generative AI into expressive art therapy to help children and families express emotions and thoughts. While the study demonstrated the potential of AI as a tool to enhance participants' creative potential, it also highlighted the risks, particularly regarding safety due to the uncertainties in generative AI that could lead to inappropriate or unsuitable content. To reduce the risk, the study recommended the development of more age-appropriate generative AI interfaces through collaboration among parents, children, and therapists.}\rev{ 
 A study by Yilma et al.~\cite{yilma2024artful} introduced AI-driven VA RecSys engines 
to recommend artworks tailored to the emotional and psychological needs of PICS patients. 
These systems leverage text-based, image-based, and multimodal methods to analyze and recommend art. 
Text-based VA RecSys employ Natural Language Processing (NLP) techniques, 
such as LDA~\cite{blei2003latent} and BERT~\cite{devlin2018bert}, 
to interpret textual descriptions of art. 
Image-based VA RecSys, which analyze visual features like color and composition, 
proved more effective in aligning recommendations with therapeutic goals. 
These systems prioritized calming and uplifting artworks, 
which are essential for emotional well-being during recovery~\cite{yilma2024artful}. 
Multimodal VA RecSys, such as BLIP~\cite{yilma2023together}, integrating textual and visual inputs, 
offered a balanced approach that improved recommendation accuracy. 
}

\rev{
These findings highlight the importance of designing AI systems that incorporate healing elements, 
such as sensory pleasure and engagement, to maximize their therapeutic impact. 
However, those results also underscore a critical risk; 
some of the AI-based approaches generated recommendations of paintings 
that included imagery of ruins, destruction, and darker aesthetics—elements, 
which are particularly dangerous and inappropriate for therapeutic settings. 
These outcomes highlight that while AI shows considerable promise in supporting art therapy, it cannot be relied upon in isolation. 
Human intervention is pivotal to ensure that the therapeutic content is both safe and effective. 
To address this challenge, the present study explores human-AI collaboration 
by proposing a collaborative approach that integrates both therapist expertise and AI capabilities for art therapy in PICS interventions.
}

\section{Method: Human-in-the-Loop VA RecSys for PICS intervention}
\label{sec: method}
\rev{
AI-driven VA RecSys can analyze patient data to suggest artworks that might resonate with their emotional and psychological states. However, without human oversight, these systems might miss the subtle understanding needed to ensure that the recommendations are not only relevant but also therapeutically beneficial~\cite{yilma2024artful}. While these collaborative frameworks present interesting research avenues, our work focuses on a human-AI collaborative approach. Particularly in the context of art therapy by visual exposure, such collaboration becomes vital, as it ensures that human therapists can oversee and refine AI recommendations, making them more effective, risk-free and aligned with patient needs. Therefore, in this work, we propose a HITL approach that leverages the strengths of both human and AI inputs to enhance the effectiveness of art therapy for PICS patients.
}

We adopted the two best-performing VA RecSys architectures previously validated in the context of art therapy~\cite{yilma2024artful}. 
Specifically, we developed a visual-only VA RecSys model using the pre-trained ResNet-50 architecture~\cite{he2016deep}, and a multimodal (visual + tex VA RecSys using the pre-trained BLIP model~\cite{li2022blip}. The following section outlines how the VA RecSys pipeline operates.

\begin{figure*}[!ht]
\centering
\centering
\includegraphics[width= 0.9\textwidth]{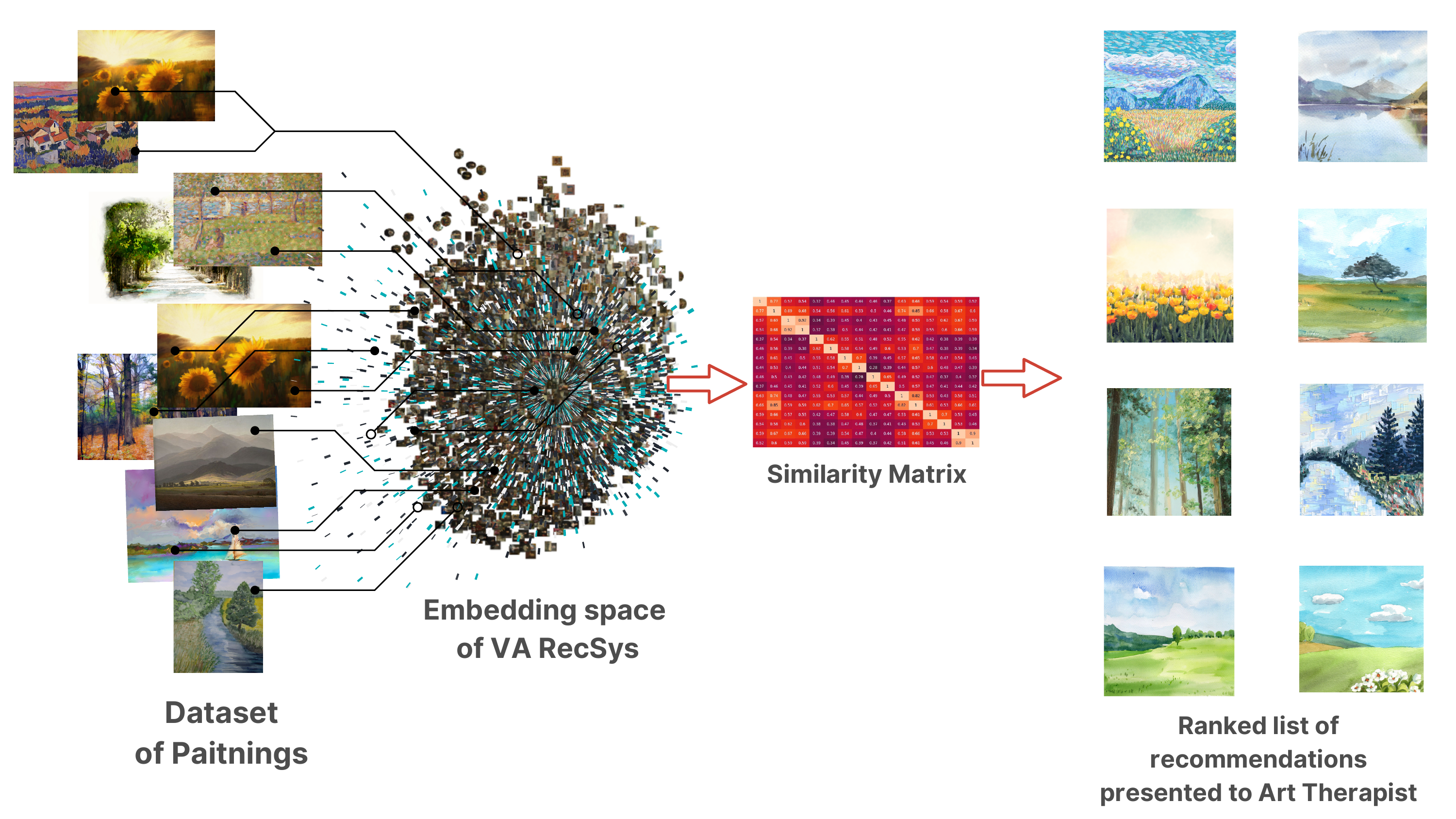}
\caption{
    Diagram of the VA RecSys pipeline, from automatic feature extraction to automatic painting selection.
}
\Description{
    Diagram of the VA RecSys pipeline, from automatic feature extraction to automatic painting selection. The expert (\rev{Art Therapist}) then curates these painting and finally delivers them to the PICS patients.
}
\label{fig:VA Recsys-pipeline}
\end{figure*}

Let $P = \{p_1, p_2, \dots, p_m\}$ represent the set of paintings, and
and $\mathcal{P} = \{\vp_1, \vp_2, \dots, \vp_m\}$ represent the associated embeddings of each painting as computed by either ResNet-50 or BLIP. After the embeddings (latent feature vectors) for the dataset are learned using these models, we calculate the similarity matrix for all the paintings $\mathbf{A}$. The preference of a patient user $u$ is then modeled by computing a ranking score for the paintings in the dataset according to their therapeutic relevance.  Specifically, their similarity to the painting $p_j$ that the user identified as supportive of their recovery. The predicted score $S^u(p_i)$ the user would give to each painting in the collection $P$
is calculated as: 
\begin{equation}\label{eq:user-score}
    S^u(p_i) = d(\vp_i, \vp_j)
\end{equation}
where $d(\vp_i, \vp_j)$ is the cosine similarity between embeddings of paintings $p_i$ and $p_j$ in the computed similarity matrix.   
Once the scoring procedure is complete, the paintings are sorted and the $r$ most similar paintings constitute a ranked recommendation list. This list is then presented to the experts (\rev{Art therapists}) for validation. Experts have the opportunity to filter, modify, or regenerate the recommendations. Once the experts are satisfied with the selection, the curated list of paintings is embedded into the guided therapy sessions. This process is aimed at not only saving time by narrowing down thousands of potential options but also ensuring that the final therapeutic experience is both personalized and rooted in expert knowledge.

\section{Materials}

We followed the setup proposed by \citet{yilma2024artful}, aimed at facilitating study comparisons. 
For preference elicitation, a set of 18 nature-themed paintings from the WikiArt database\footnote{\url{https://www.wikiart.org}} were used, 
designed to evoke positive emotional responses, such as relaxation, cheerfulness, 
and awe-inspiring. These paintings were selected by a panel of experts in affective psychology, environmental psychology, and healthcare design, and were evaluated with 186 participants through a pre-study approved by the Ethics Review Panel of the University of Twente. This evaluation process led to the final selection of three paintings 
that were most positively perceived by participants.

For generating painting recommendations, we utilized a collection of 2,368 paintings from the National Gallery, London, available under a Creative Commons (CC) license through the CrossCult Knowledge Base.\footnote{\url{https://www.crosscult.lu}} 
This dataset, with its comprehensive metadata, provided a robust foundation for feature extraction and analysis.
In terms of technical implementation, painting features were extracted using two pre-trained models: ResNet-50 and BLIP. 
The ResNet-50 model was employed to extract convolutional feature maps, capturing the visual essence of the paintings. 
The BLIP model, which integrates visual and textual information, was used to create multimodal embeddings for each painting. These models were then used to create two lists with 200 painting recommendations each, 
which were then subject to expert validation, 
ensuring their relevance and effectiveness in therapeutic settings.

\subsection{Expert evaluation}

To understand the impact of HITL approaches in guided art therapy from the perspective of an expert, 
we conducted a simple experiment focusing on two key factors: efficiency and appropriateness. 
Efficiency was evaluated quantitatively, 
by assessing whether a HITL approach could reduce the time spent
by the expert to select appropriate nature paintings for art therapy.
Appropriateness, on the other hand, was evaluated qualitatively, 
by observing whether the expert's intervention was necessary 
to safeguard the quality and suitability of the paintings recommended by each HITL approach.

The expert’s role was twofold: to confirm that the paintings featured the correct distinctive qualities, 
making them sufficiently relevant to the original pieces,
and to avoid selecting paintings with potentially negative effects on viewers. 
The expert first reviewed the entire collection of paintings to become familiar with it. 
Then, the selection process for the three groups was as follows (\autoref{fig:process}):

\begin{figure*}[!ht]
\centering
\includegraphics[width= 0.9\textwidth]{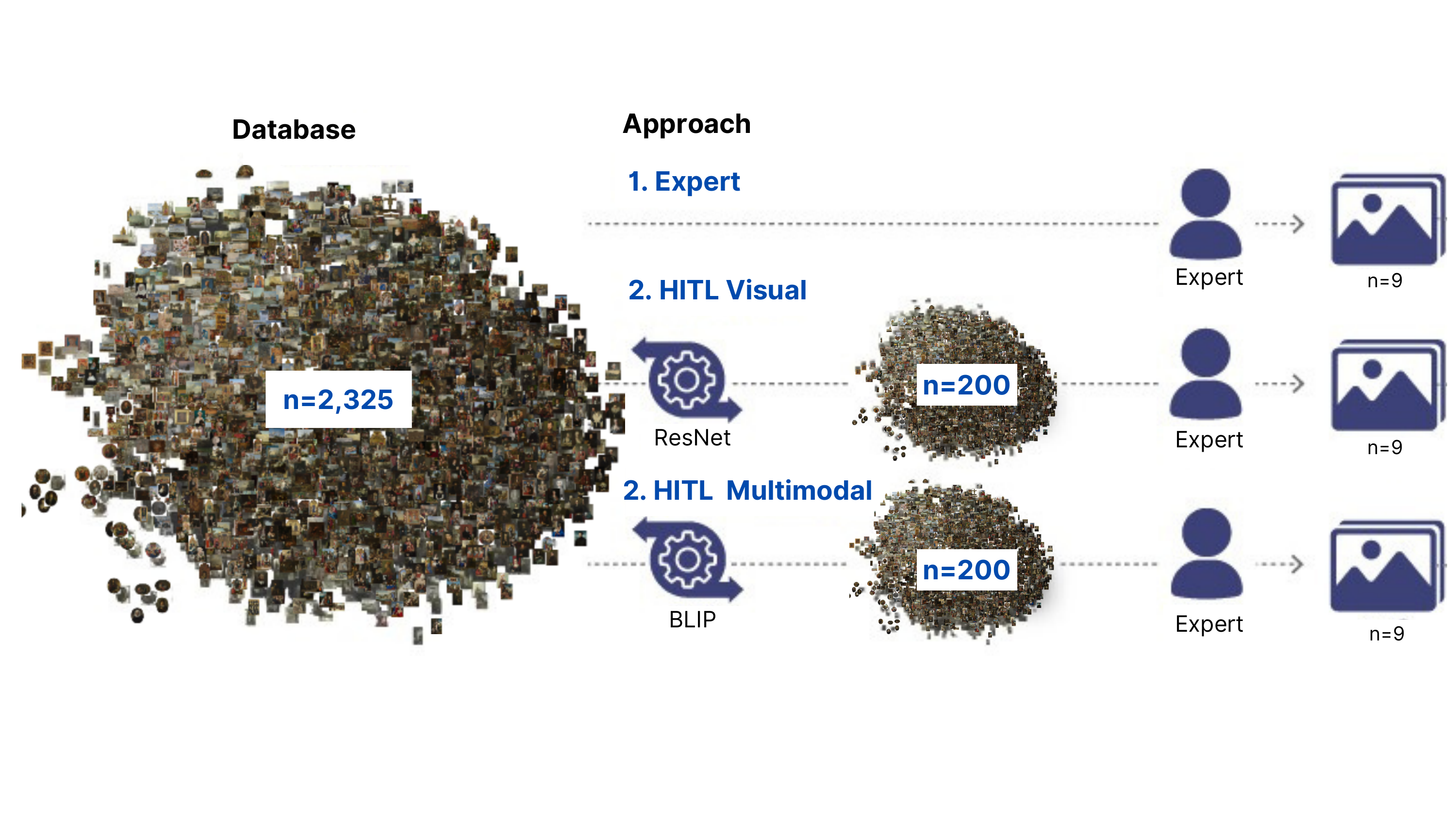}
\caption{
    Overview of the painting selection process using three different approaches: Expert, HITL Visual, and HITL Multimodal. 
}
\Description{
    Overview of the painting selection process using three different approaches. Art Therapists have to go through the whole collection of paintings, often in a random order, whereas HITL approaches provide the expert with a ranked and shorter selection, thereby allowing them to save valuable time.
}
\label{fig:process}
\end{figure*}

\begin{itemize}
    \item \textbf{\textit{Expert selection (Expert):}} Considering both the semantic and stylistic qualities of the three original paintings, the expert selected three additional paintings for each, from a dataset of 2,325 paintings, resulting in a total of nine paintings.
    
    \item \textbf{\textit{Human-AI collaboration (HITL Visual, HITL Multimodal)}}: Based on the three original paintings, both HITL approaches provided a top-200 ranked list of recommendations. From there, the expert selected three paintings most relevant to each of the three original paintings, again resulting in a total of nine paintings per HITL approach.
\end{itemize}

The time taken (in minutes) for the expert to choose the three most relevant paintings for each original painting was recorded. 
Through these three approaches, a total of 27 paintings were selected from the original pool of 2,325 for the study.

\subsection{Results}

\autoref{tab:expert-timing} provides the results for the time analysis.
The results showed that using the HITL approach with both ResNet-50 and BLIP 
reduced the time the expert spent selecting paintings by more than 50\%, 
from 37 minutes in total to 15 minutes (ResNet-50) and 17 minutes (BLIP).
The differences between both HITL approaches 
and manual work are statistically significant:
$F(2,6) = 57.56, p < .001, \eta_p^2 = 0.95$.
A pairwise $t$-test (Bonferroni-Holm corrected) 
revealed no statistically significant difference
between both HITL approaches ($p=.462$).

In terms of content relevance, the expert noted that the HITL approaches 
helped them find more appropriate paintings.
``Starting with the top-200 paintings instead of over 2,000 was very helpful. With the larger pool, I had to sift through many irrelevant paintings, which became tiring. Among the top-200 recommendations, I found a few more relevant pieces compared to starting with the full set of 2,000. These might have been overlooked among the irrelevant options, so reducing the noise helped me focus on selecting better ones.''

However, despite the efficiency improvements, the expert expressed concerns regarding content appropriateness. 
``Even within the top-50 recommendations from both VA RecSys engines, there were still irrelevant paintings, such as a portrait in shadows or dark indoor environments devoid of nature. Additionally, some paintings featured content or styles that could be emotionally distressing for patients, such as war imagery, violent storms, or scenes of injury and death''
These issues highlight the ongoing and important need for human involvement in the painting selection process, reinforcing the idea that HITL approaches are preferred over fully automated approaches.

\begin{table}[!h]
\centering
\small
\caption{
    Comparison of the time (in minutes) 
    taken by an expert to find relevant paintings 
    using different approaches.
}
\label{tab:expert-timing}
\begin{tabular}{l *5r}
\toprule
\textbf{Approach} & \textbf{Painting 1} & \textbf{Painting 2} & \textbf{Painting 3} & \textbf{Total time} & \textbf{Mean $\pm$ SD time} \\
\midrule
Expert          & 11.27 & 12.08 & 14.15 & 37.50 & 12.5 $\pm$ 1.21 \\
HITL Visual     &  5.17 &  5.08 &  5.23 & 15.48 & \textbf{5.16 $\pm$ 0.07} \\
HITL Multimodal &  5.33 &  5.47 &  6.57 & 17.27 & 5.75 $\pm$ 0.62 \\
\bottomrule
\end{tabular}
\end{table}

\section{User study}

The main goal of our evaluation was to understand the user's perception towards the quality of our VA recommendation strategies for PICS rehabilitation therapy, and ultimately to assess their efficacy in supporting the healing journey of PICS survivors. 
We conducted a large-scale user study that was approved by the Ethics Review Panel of the University of Twente.

\subsection{Apparatus}

We designed an online guided therapy survey using Google Forms\footnote{\url{https://www.google.com/forms/about/}} that first elicited the preferences of participants by providing them with a set of paintings and asking them to select one that resonates with their healing journey. Then participants were taken through a guided therapy session by using three paintings that were recommended based on their initial choice.

\subsection{Participants}

We recruited a balanced pool of $N=150$ participants
via the Prolific crowdsourcing platform.\footnote{\url{https://www.prolific.co}} 
The screening criteria were:
(i)~being active in the platform in the last 90 days,
(ii)~100\% approval rate in previous Prolific studies,
(iii)~being proficient in English,
and (iv)~having been hospitalized post-COVID (as the likelihood of having developed PICS is higher).
The latter screening criterion was phrased as
“I have been officially diagnosed with COVID-19 (tested by a licensed medical professional), and was treated in a hospital.”

Our recruited participants (74 female, 75 male, 1 prefer not to say) 
were aged 31.7 years (SD=10.7) and could complete the study only once. 
Most of them lived in UK (66 participants) or USA (55).
Most participants mentioned having been through general ward or a medical/surgical unit (105), 
or in ICU (69) or in an Emergency Room (61).
For most participants, the duration of their stay in a hospital was 
less than a week (128) or between 1 to 2 weeks (76).
Twenty-nine participants stayed in the hospital for almost one month,
and 15 participants stayed between 1 and 2 months.

The study took a median time of 24\,min to complete and participants were paid an equivalent hourly wage of \$12/h.  
We also administered the Patient Health Questionnaire-4 (PHQ-4)~\cite{lowe20104} 
to collect signs of psychological symptoms related to PICS: anxiety and depression. 
Most participants declared to suffer from anxiety (132) and/or depression (119), 
indicating the presence of psychological components related to PICS symptoms. 

\subsection{Design}

We deployed two HITL engines together with the Expert recommendations: 
`HITL Visual' (ResNet-based) and `HITL Multimodal' (BLIP-based) VA RecSys. 
Each participant was only exposed to the recommendations generated 
by one of the three groups (between-subjects design) and then went through guided art therapy. 
Each group comprised 50 participants.

\subsection{Procedure}

We first assessed baseline and post-test affective states using two different measures: 
a Pick-A-Mood (PAM) tool~\cite{desmet2016mood} for quantifying mood 
and the short version of the Positive and Negative Affect Schedule (PANAS) scale~\cite{watson1988development} 
for quantifying emotions. \revTwo{Two different affective measurement tools—PAM for mood state and PANAS for emotional state—assess the effect of temporal affective state enhancement resulting from the intervention. This can indicate the effect of the intervention on addressing psychological aspects of PICS, for instance, related to anxiety and depression.}

In order to reduce the cognitive load on participants and to be more efficient, 
to assess the affective state of participants before and after guided art therapy,
we selected 10 items that are relevant to PICS (for negative emotions) and patient well-being (for positive emotions). 
Particularly we consider 5 positive and 5 negative items (instead of 10 positive and 10 negative items) 
together with a neutral item.

Guided art therapy involves asking questions that encourage participants to engage with the paintings, 
such as \textit{``Imagine yourself entering the painting and exploring it. How did you feel while spending time in this painting?''} 
Participants were prompted to reflect on their experience and describe it in three to four sentences. 
Participants also rated the provided painting recommendations in a 5-point Likert scale. 
Our dependent variables are widely accepted proxies of recommendation quality~\cite{pu2011user}:
\begin{description}
    \item[Accuracy:] The paintings match my personal preferences and interests.
    \item[Diversity:] The paintings are diverse.
    \item[Novelty:] I discovered paintings I did not know before.
    \item[Serendipity:] I found surprisingly interesting paintings.
\end{description}

We also collected two dependent variables that inform to what extent the recommended paintings 
contributed to a sense of immersion and engagement:
\begin{description}
    \item[Immersion:] How much do the recommended paintings contribute to your sense of immersion, 
        making you feel deeply involved or absorbed in the artwork?
    \item[Engagement:] To what extent do the recommended paintings contribute to your feeling of engagement, 
        capturing your attention and generating a sense of involvement or interest?
\end{description}

\subsection{Results}

We investigated whether there were differences between the three recommendation groups (i.e. Expert, HITL Visual, and HITL Multimodal).
We used a linear mixed-effects (LME) model where each dependent variable is explained by each recommendation group.
Participants are considered random effects.
An LME model is appropriate here because the dependent variables are discrete and have a natural order.

We fitted the LME models (one model per dependent variable) 
and computed the estimated marginal means for specified factors.
We then ran pairwise comparisons (also known as \emph{contrasts})
with Bonferroni-Holm correction to guard against multiple comparisons.

\subsubsection{Analysis of recommendation quality measures}

\begin{figure*}[!ht]
    \centering
    \def\w{0.3\linewidth}
    
    \includegraphics[width=\w]{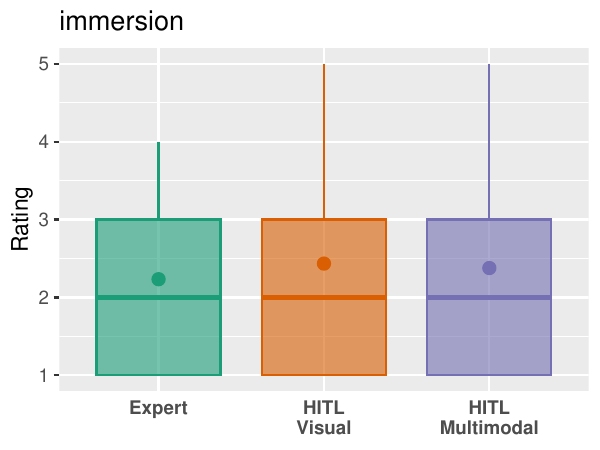} \hfill
    \includegraphics[width=\w]{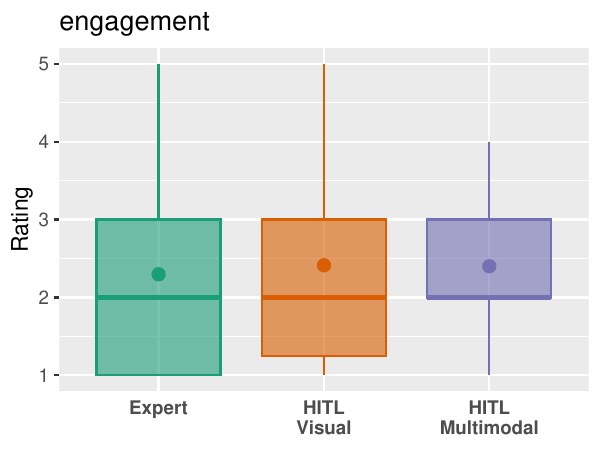} \hfill
    \includegraphics[width=\w]{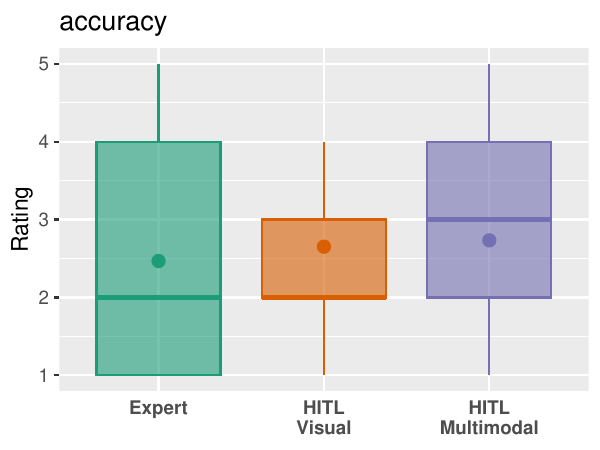} \\
    \includegraphics[width=\w]{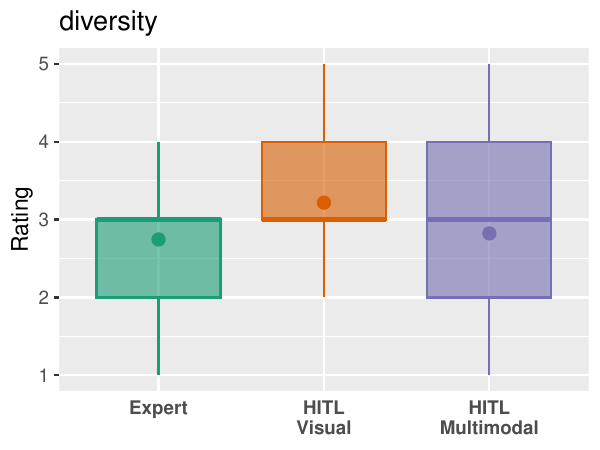} \hfill
    \includegraphics[width=\w]{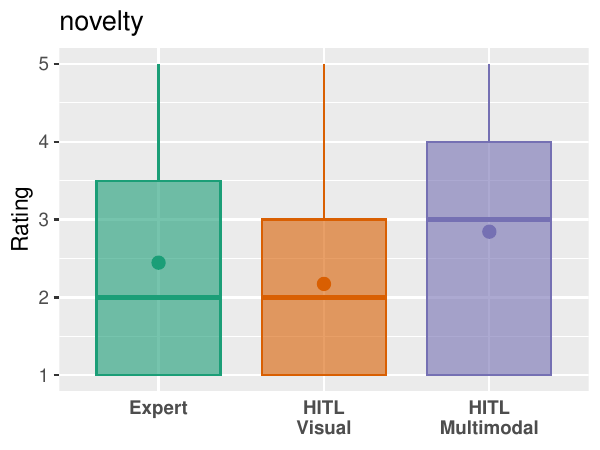} \hfill
    \includegraphics[width=\w]{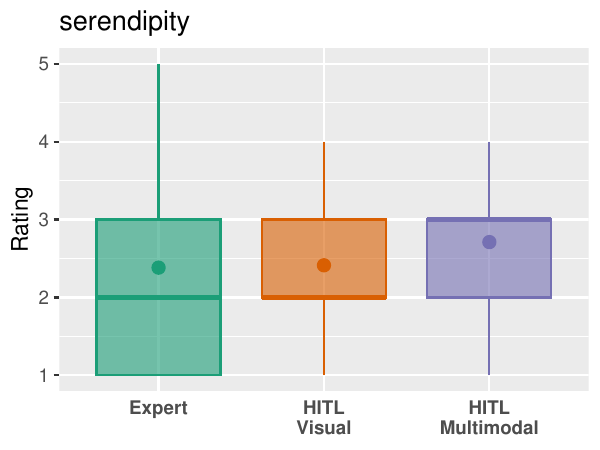} 

    \caption{
        Distribution (box plots) of user ratings for the user-centric dependent variables of recommendation quality. Dots denote mean values.
    }
    \Description{
        Statistical plots of user ratings according to each of our dependent variables: immersion, engagement, accuracy, diversity, novelty, and serendipity. All RecSys groups were found to perform similarly, comparable to Expert-level performance.
    }
    \label{fig:prolific-ratings-overall}
\end{figure*}

\autoref{fig:prolific-ratings-overall} shows the distributions of user ratings
for the user-centric dependent variables of recommendation quality.
We only found statistically significant differences
between HITL Visual and HITL Multimodal engines in term of novelty:
$\chi^2(129) = -2.447, p=.0472, r=0.210$,
indicating that HITL Multimodal was preferred over HITL Visual.
All other differences between groups were not statistically significant for any of the other dependent variables, 
with small to moderate effect sizes.

\subsubsection{Analysis of changes in mood}

\begin{figure*}[!ht]
    \centering
    \includegraphics[width=\linewidth]{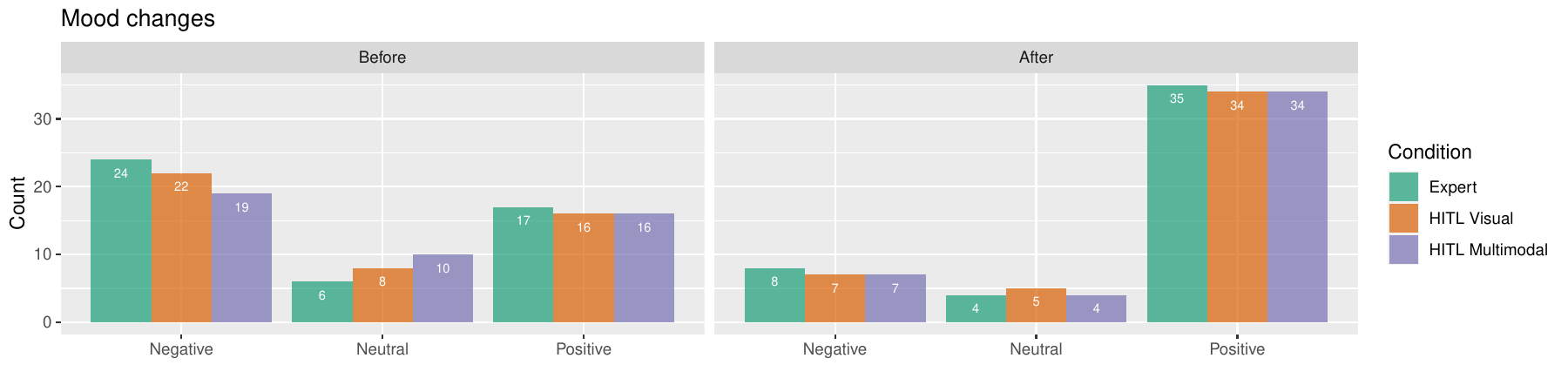}
    \caption{
        Mood improvement comparison before and after art therapy.
    }
    \Description{
        Bar plot showing mood scores according to three groups: Expert vs. HITL Visual vs. HITL Multimodal recommendations. All groups showed a comparable positive improvement after Art therapy.
    }
    \label{fig:mood_improvement}
\end{figure*}

\autoref{fig:mood_improvement} shows the mood changes before and after therapy,
for the three recommendation groups we have considered in our study. 
In all three groups, a mood enhancement effect of guided art therapy was observed. 
When comparing it to the baseline where the majority of participants were in a negative mood (47.1\%), 
after guided art therapy the majority of participants reported being in a positive mood (72.6\%), 
with only a minority remaining in a negative (17.1\%) or neutral (10.1\%) mood.
The differences between approaches were not statistically significant:
$\chi^2_{(2, N=38)} = 0.68, p = .710, \phi = 0.134$.

\begin{figure*}[!ht]
    \centering
    \includegraphics[width=\linewidth]{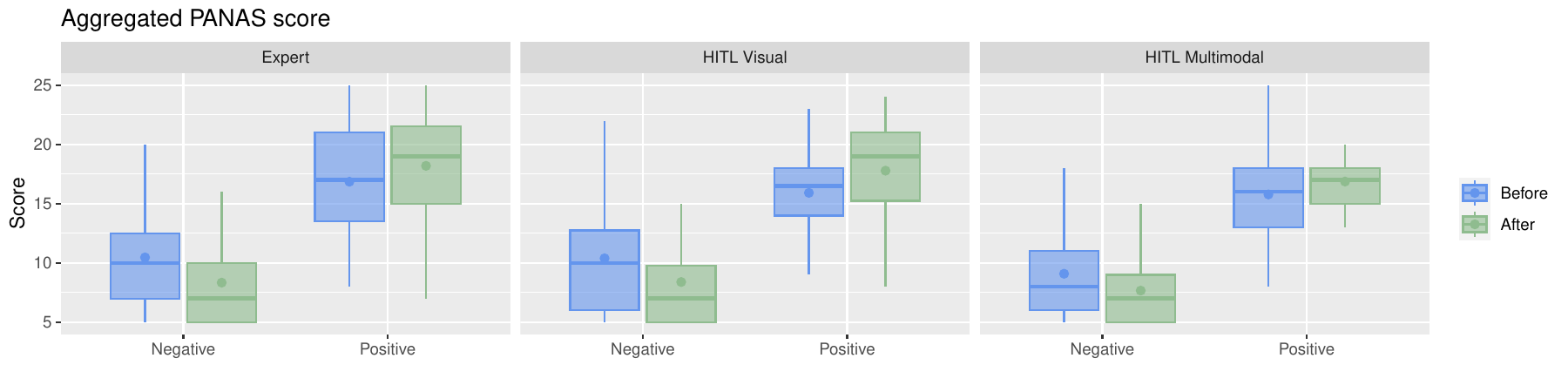}
    \caption{
        Emotion score changes according to the Positive Affect Negative Affect Schedule (PANAS) scale.
    }
    \Description{
        Statistical plots of the aggregated PANAS score according to each dimension. All RecSys groups were found to perform similarly, comparable to Expert-level performance.
    }
    \label{fig:prolific-ratings-diff}
\end{figure*}

\autoref{fig:prolific-ratings-diff} shows the change in scores after therapy,
aggregated according to the ten items of the PANAS scale.
Differences between groups were not statistically significant in any case, with small to moderate effect sizes.
According to an item-independent analysis,
the largest effect sizes were observed in terms of the `afraid' item,
when comparing Expert recommendations against Visual ($r=0.181$) and Multimodal ($r=0.175$) recommendations,
followed by the `scared' item when comparing Expert and Visual recommendations ($r=0.122$). 
Upon further examination, users in the Visual group did not change
their scores for the `afraid' item (the median difference is 0).
This was also the case for users in both the Visual and Multimodal group
with regards to the `scared' item.

\begin{figure*}[!ht]
    \centering
    \def\w{0.23\linewidth}
    
    \includegraphics[width=\w]{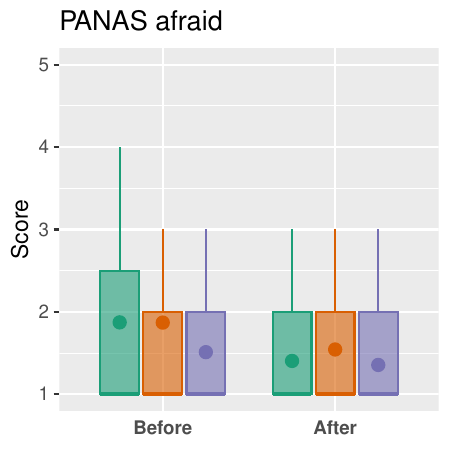}
    \includegraphics[width=\w]{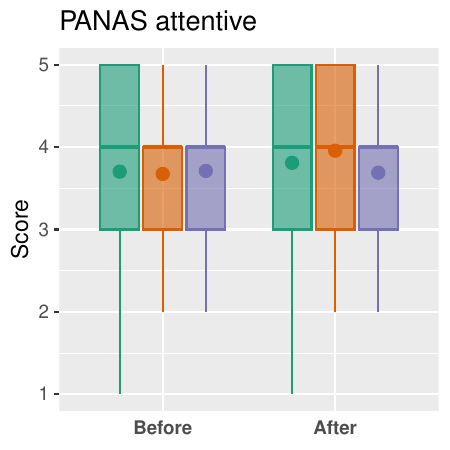}
    \includegraphics[width=\w]{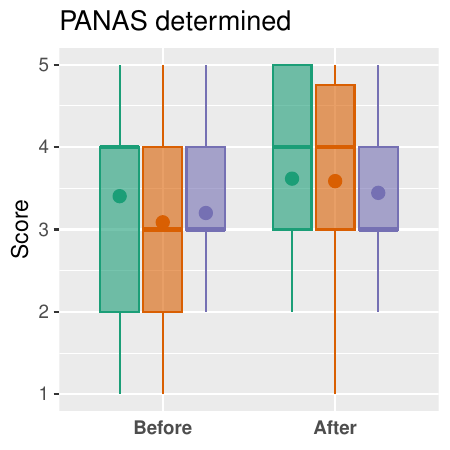}
    \includegraphics[width=\w]{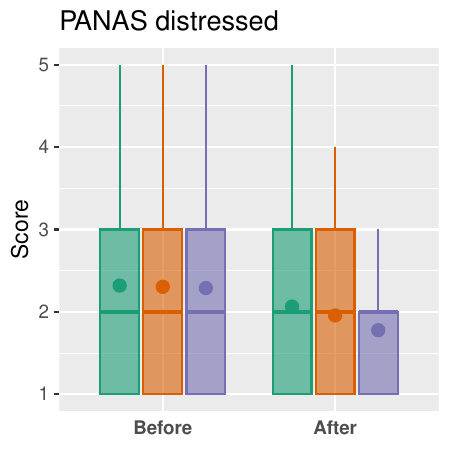}
    \includegraphics[width=\w]{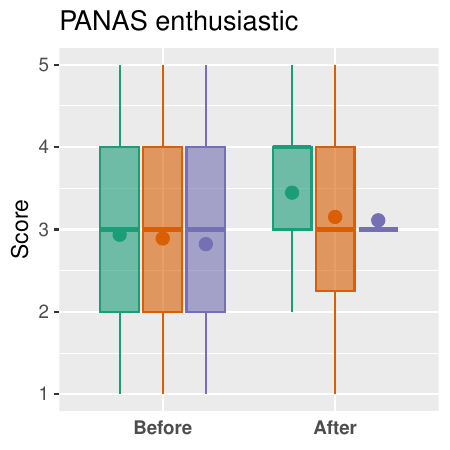}
    \includegraphics[width=\w]{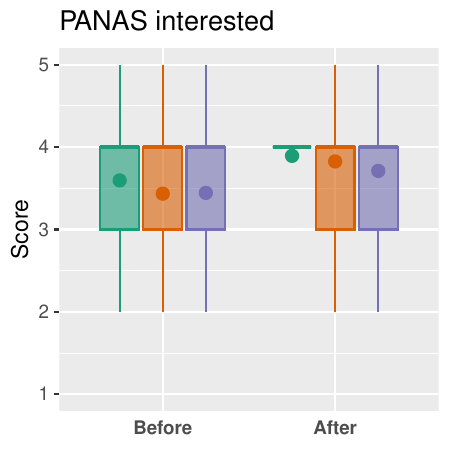}
    \includegraphics[width=\w]{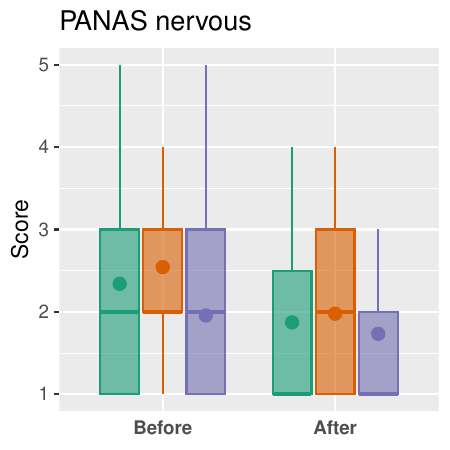}
    \includegraphics[width=\w]{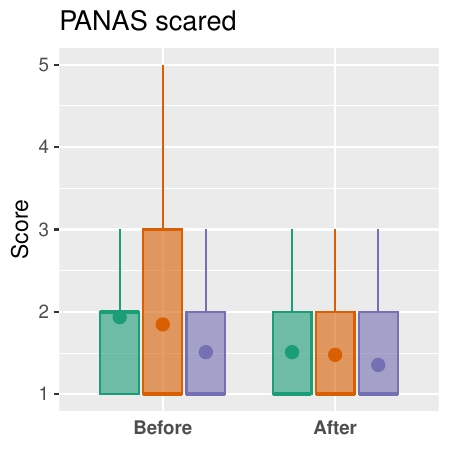}
    \includegraphics[width=\w]{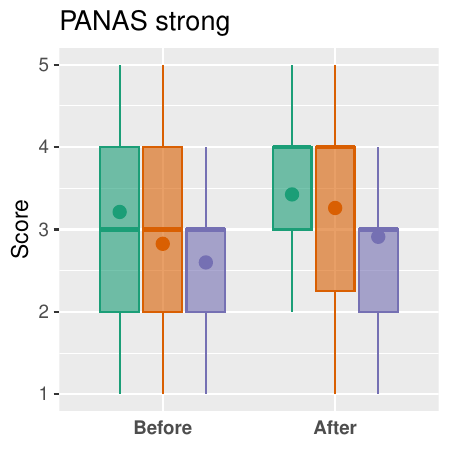}
    \includegraphics[width=\w]{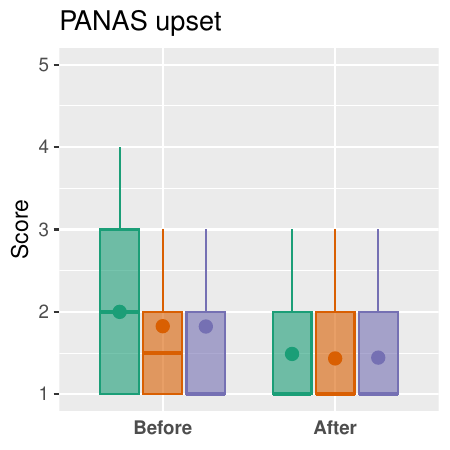}
    \phantom{\includegraphics[width=\w]{plots/2025/panas_upset.pdf}}
    \phantom{\includegraphics[width=\w]{plots/2025/panas_upset.pdf}}

    \caption{
        Individual Positive Affect Negative Affect Schedule (PANAS) scores per dimension.
    }
    \Description{
        Statistical plots of PANAS scores (short version) according to: afraid, attentive, determined, distressed, enthusiastic, interested, nervous, scared, strong, upset. All RecSys groups were found to perform similarly, comparable to Expert-level performance.
    }
    \label{fig:panas-individual}
\end{figure*}

\subsubsection{Analysis of user reflections}

We performed sentiment analysis on participants' reflections to obtain a better understanding of their experiences. 
For this, we utilized the \texttt{bert-large-uncased-sst2} model, 
a pre-trained Transformer-based sentiment analysis model 
from the Hugging Face Transformers library.\footnote{\url{https://huggingface.co/models}} 
This model, which builds on \texttt{bert-large-uncased}, 
has been fine-tuned using the Stanford Sentiment Treebank v2 (SST2)\footnote{\url{https://nlp.stanford.edu/sentiment}} 
as part of the General Language Understanding Evaluation (GLUE) benchmark.\footnote{\url{https://gluebenchmark.com}}
Due to its extensive training and versatility, this model excels in various NLP tasks, including sentiment analysis. 

\begin{figure*}[!ht]
    \centering

    \def\w{\textwidth}
    \includegraphics[width=\w]{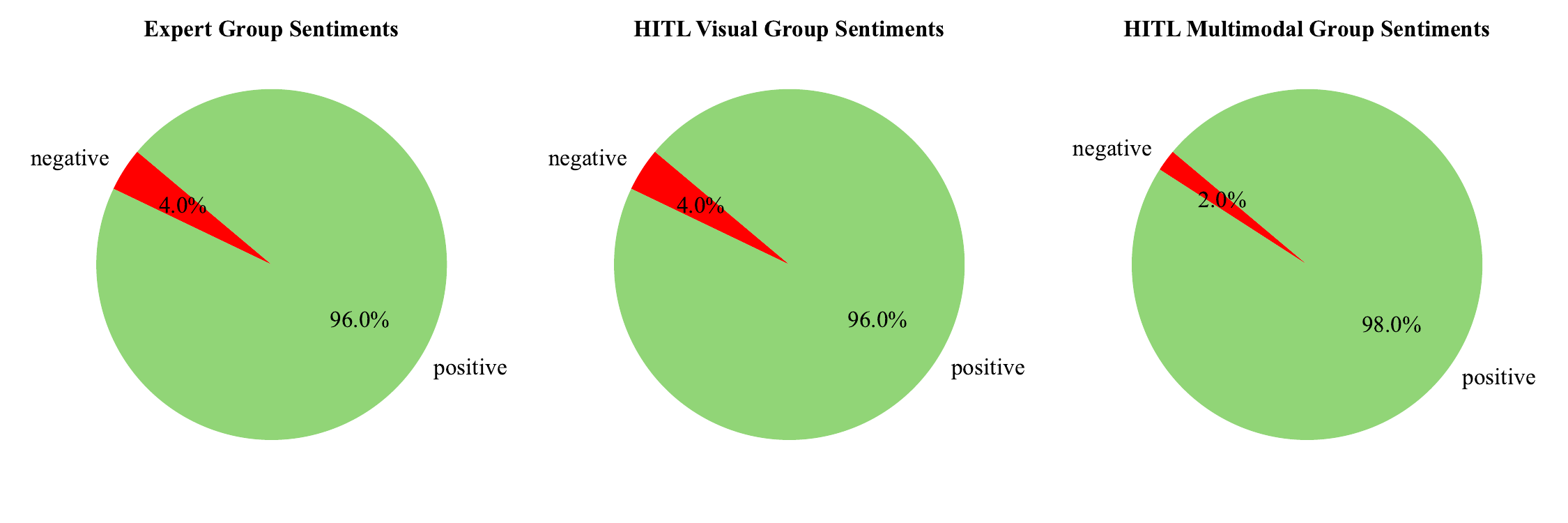}

    \caption{
        Sentiment analysis of user reflections per group.
    }
    \Description{
        A pie chart based on Text-based sentiment analysis of the user reflections showing the proportion of positive and negative sentiments of users
    }
    \label{fig:sentiment}
\end{figure*}

\begin{figure*}[!ht]
    \centering

    \def\w{\textwidth}
    \includegraphics[width=\w]{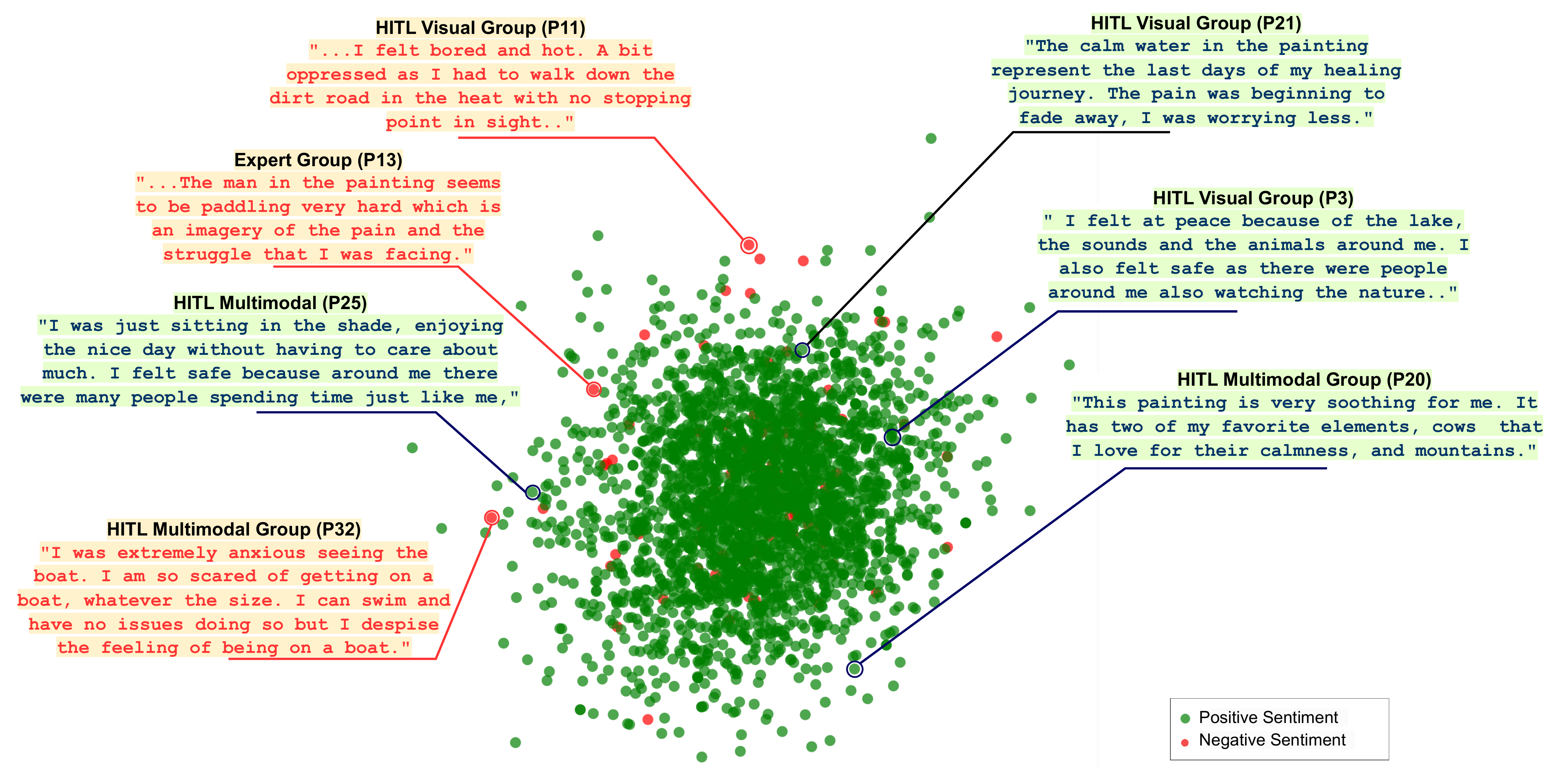}

    \caption{
        Non-linear projection (t-SNE) of sentence embeddings from user reflections.
    }
    \Description{
        Plot showing a 2D projection of paintings according to the features extracted by the BERT language model. A cloud of dots is shown, where each dot represents a user sentiment. A few dots have annotations that depict the sentiment expressed by a user.
    }
    \label{fig:embeddings_sentiment}
\end{figure*}

The outcome of our sentiment analysis, shown in \autoref{fig:sentiment}, 
reveals predominantly positive sentiments among participants regarding their engagement with the recommended paintings. \rev{Negative responses were primarily due to over- or understimulation of the paintings, or from negative memories (e.g., inducing anxiety) triggered by the paintings. }
Interestingly, a subtle trend emerged among the Expert group and the HITL Visual group, 
where 4\% of the sentences reflected negative sentiments. 
In contrast, the HITL Multimodal group exhibited a slightly lower negativity rate of 2\%. 
This pattern mirrors the results observed in the mood change PANAS scores and the measures of recommendation quality. 
\autoref{fig:embeddings_sentiment} shows some sample sentences from the positive and negative reflections of participants 
per group on a 2D projection map of all reflection sentences 
using the non-linear projection t-SNE algorithm~\cite{Maaten:2008:tSNE}.
 
Although the sentiment analysis provided useful insights, 
we acknowledge that our pre-trained sentiment analysis model based on the general-purpose language model BERT, 
may not fully capture nuanced sentiments in the specific context of healing. 
As the effectiveness of AI-recommendations is inherently tied to the richness of data (including richness in terms of subjective interpretations of artworks), another aim of our work was to extend the findings of \citet{yilma2024artful} by verifying and potentially identifying additional themes over and above the seven themes defined in their study 
(hope, rejuvenation, engagement, safety, sensory pleasure, relevance, and personal preference). 
To this end, a random selection of participants responses to paintings 
(comprising 40\% of all participant responses) was coded. 

In addition to confirming the previous theme categorization by \citet{yilma2024artful},
 three additional themes were identified, 
which were clearly connected to positive affective states conducive to recovery: 
\textit{Togetherness vs Solitude,} \textit{Awe and Feeling small} and \textit{Escape and Refuge}. 
These themes were found in all three groups. 
We elaborate on these themes with example quotes. 

\textit{Togetherness vs Solitude.} 
This code represents descriptions in which participants enjoy being together with others but at the same time indicate a preference for keeping social interactions at a distance as they might be too taxing or overwhelming:
“In this painting, I was reading a book and looking at people. The breeze from the lake was hitting my face and the smell of the grass was potent. I felt a little anxious because there were some people around, but it was nice to hear the conversations.”
The following description likewise hints at a balance between needs for solitude and togetherness (and the imagined capability to switch between the two): 
“I imagine being with my family a bit apart from each other looking out at the water 
disconnected from each other’s thoughts and feelings, 
but all I have to do is point them to the rainbow to the left 
and bring us together and connect to the simple yet most amazing wonder, 
the rainbow represents color, diversity and togetherness”. 

\textit{Awe.} 
Awe, a sense of feeling small in the presence of something greater than the self~\cite{keltner2003approaching}, 
regularly surfaced in the positive experiences induced by the paintings as illustrated in the following quotes stressing vastness or the grandeur of a landscape or natural elements: 
“In this painting, I captured life's relentless surge in this maritime realm serene yet unsettling, 
its darkness and monotony drawing me in. I immerse myself in nature's embrace, 
the boundless sky mirrored in the lake's depths, a vivid testament to its grandeur.” 
and “In this painting I was sitting outside, looking at the houses around me, 
as well as the giant trees surrounding the area. 
Wow, I thought to myself. These tall trees must have been here for hundreds of years. 
Look at what Mother Earth has given us.” 

\textit{Escape and Refuge.} 
The last additional theme is represented by narratives in which participants describe a sense of refuge or an escape from their life as it is: 
“This experience serves as a refuge, a momentary escape from life's demands, 
helping me reconnect with the core of my being and find solace in the beauty of the world around me.”
A similar sense of refuge in nature and retreat from everyday life can be found in the following quote where calmness and tranquillity prevail, and the city is far away:
“In this painting I just imagined myself relaxing sitting on the grass watching and enjoying the sunset on the horizon, observing the tranquillity of the sky and the rainbow after a storm, nothing matters at this moment, just enjoying the beauty of the sky, the calm river and the city that can be seen in the distance.”

\autoref{tab:themes} provides an overview of the themes along with their descriptions and example quotes:
Seven themes are aligned with the work of \citet{yilma2024artful} and three themes are newly discovered by our study.

\begin{table*}[htbp]
\small
\begin{tabularx}{\textwidth}{p{0.15\linewidth}X<{\arraybackslash}p{0.55\linewidth}X}
\toprule
\textbf{Themes}  & \textbf{Theme description} & \textbf{Example quotes} \\
\midrule
Hope and Purpose  & Drawing one's attention toward more positive prospects, reminding them of hope and purposefulness. & \texttt{“I sit next to the water, listening to the sound of nature as the birds sing sweet songs. I analyze what I have become and what the future holds for me. Going forward with my chin up, I am stronger than before and not afraid of anything, unlike before when I made mistakes.”} \textbf{P20}
\\
\midrule
Rejuvenation & Supporting one to feel recharged through a sense of being carefree, calm, and relaxed. & \texttt{“I felt happy because I was calm and nothing was bothering me in any way. I felt relaxed. I took a journey to the clouds and saw everything in my vision. The journey of relaxing and thinking things through made me realize that, despite life's challenges, I now see myself as a bigger person.”} \textbf{P20} \\
\midrule
Engagement  & Supporting one to immerse into the visual art by triggering one’s attention and interest. & \texttt{“I like the way they layer their colors together to account for the shade the clouds create, and the attention to detail keeps me focused on the painting. This allows all my worries to fade away and lets me be immersed in the experience.”} \textbf{P11} \\
\midrule
Safety  & Promoting a sense of safety through elements that signal a safe and familiar environment. & \texttt{“In this painting, I was walking around, exploring big trees, and decided to sit on a bench to relax and calm my thoughts. It was nice and warm, and I felt very secure and protected.”} \textbf{P14}
\\
\midrule
Sensory Pleasure & Providing pleasure through rich sensory stimulation that either directly comes from visual art or is derived from memory through visual triggers. & \texttt{“... I could even smell the fresh air and feel the ground at my feet. I imagined myself walking along the path in complete silence. Only thing I could hear was the wind and maybe some animals like a cow.”} \textbf{P17} 
\\
\midrule
Relevance & Presenting subject matter relevant to one’s specific situation that can stimulate memories or constructive reflections. & \texttt{“This moment reminds me of the carefree childhood I spent in the countryside. Thanks to these memories, I can calm down and know that I can handle everything.”} \textbf{P7}\\
\midrule
Personal Preference & Increasing pleasure with visual stimulation that meets one’s preference. & \texttt{“Many people find mountains to be a source of inspiration, just like myself.”} \textbf{P32} 
\\
\midrule
Togetherness vs Solitude  & Balancing the need for connectedness without being overwhelmed during recovery. & \texttt{“...The loneliness of a journey that begins with the illness that you yourself must overcome, but that has an end. There in the mountains, once climbed down, health awaits you ... Meanwhile, some passenger joins you, the rest of the patients with your same pathology whose experience makes the road easier.”} \textbf{P54} 
\\
\midrule
Awe  & Promoting a sense of feeling small in the presence of something greater than the self. & \texttt{“Wow, I thought to myself. These tall trees must have been here for hundreds of years. Look at what Mother Earth has given us. ... I was sitting down in a chair being thankful that my life has a meaning and the world around me has given me such positive experiences. This experience made me feel like I can move past the experience that I had previously in the ICU.”} \textbf{P16} 
\\
\midrule
Escape and Refuge   & Promoting a sense of getting away from an unpleasant or stressful situation. & \texttt{“... A place with almost no foot traffic or noise. I also imagined myself sitting down and having a peaceful lunch. It feels like a place where I can go when I need to escape from people or my worries for a while.”} \textbf{P17}
\\
\bottomrule \\
\end{tabularx}
\caption{
    Ten healing themes identified in the recommended nature paintings, theme descriptions 
    and example quotes from some participants.
}
\Description{
    Table showing some quotes from our participants, grouped by thematic areas: Hope and Purpose, Rejuvenation, Engagement, Safety, Sensory Pleasure, Relevance, Personal Preference.
}
\label{tab:themes}
\end{table*}

\newpage
\section{Discussion}


In the current trend of integrating AI throughout various stages of the care path in a broad healthcare context, the role of experts and AI, and finding effective collaboration between them, has emerged as an important topic. \rev{AI can improve intervention and reduce the workload of healthcare professionals~\cite{fiske2019your}. In the context of art therapy, AI achieves this by analyzing vast amounts of data and generating recommendations that humans might overlook. However, the use of AI in healthcare also raises various ethical concerns and risks, including data ethics such as data bias and privacy concerns, potential errors, the potential for misuse of technologies, and the exacerbation of existing health inequalities~\cite{fiske2019your}. Previous studies~\cite{fiske2019your, cross2024use} acknowledged these issues and recommend the need for, among others, supervision in the application of embodied AI as one of the solutions and consideration of long-term effects of applications based on in-depth understanding of illness and the human condition, which can be addressed by involving human healthcare professionals and specialists in the process. This resonates with the result of our study, which shows that expert involvement is crucial throughout the process; AI cannot fully replace this role, as experts (art therapists) are essential to safeguard judgment, ensure patient safety, and provide personalized and optimal care.}

Based on our findings, we can affirmatively answer the research question posed at the outset of this paper: a collaborative Human-AI approach in art therapy indeed enhances both personalization and therapeutic outcomes for PICS patients. This HITL strategy not only optimizes the selection of therapeutic art but also amplifies the effectiveness of therapy by leveraging the complementary strengths of human expertise and AI. It also gives control to the therapist, instead of relying on fully automatic recommendations. The implications of this approach are far-reaching, as we discuss in the following sections, touching on advancements in personalization, the evolving role of AI in therapeutic contexts, and the broader impact on HCI.

\subsection{Harmonizing human expertise and AI to harness HITL in guided art therapy}
\label{subsec:discuss_one}
 
A HITL approach has demonstrated a significant mood-enhancing effect in post-ICU patients, which aligns with the results of \citet{yilma2024artful}. Additionally, HITL approaches have proved to be efficient in the preparation process, reducing the time \rev{art therapists} spend selecting paintings by 50\% and improving the quality of selections by minimizing noise. We tested three sets of paintings in this study. Considering the therapy's further implementation and its potential for long-term and multiple uses, the contribution of AI in improving efficiency is very promising, especially in light of the globally increasing shortage of healthcare professionals~\cite{marc2019nursing,wang2005assessing}. 

One of the key insights from our study is the delicate balance required between automation and human judgment in therapeutic contexts. While AI offers the capability to analyze vast amounts of data and generate recommendations that might be overlooked by humans, the risk of relying solely on AI-generated content is non-trivial.  As seen from previous studies although AI-only recommendations were compelling, they carried risks, particularly in terms of appropriateness and emotional resonance.

The HITL approach mitigates these risks by placing human expertise at the center of the decision making process. Clinicians act as gatekeepers, validating, filtering, and adapting AI suggestions before they are implemented in therapy sessions. This ensures that the therapy remains tailored to the patient's individual emotional and psychological needs while benefiting from the efficiency and breadth of AI-driven recommendations. The implications for HCI are significant, suggesting that future systems should prioritize interfaces that allow seamless interaction between human and AI agents, supporting a workflow that enhances rather than overrides human judgment. However, despite these promising results, we observed some limitations in the HITL approach, notably negative responses from participants regarding the selected paintings.

\rev{As discussed, some limitations in the HITL approach relate to negative experiences, in the current study predominantly related to overstimulation or understimulation, or by negative memories triggered by the paintings. The former might be mitigated by optimizing RecSys to ensure a diverse selection of paintings with optimal arousal levels (related to formal properties of paintings such as visual complexity and color saturation), and overall positive qualities (for a higher rate of positive experiences) \cite{kim2023morning}. However, anticipating the latter is more challenging as they do not relate to formal qualities of paintings but are rather the outcome of the interaction between scenes or situations depicted and personal (traumatic) life experiences. }This underscores the importance of careful personalization, such as including well-composed pre-therapy questions about personal trauma. It also highlights the need for ongoing expert involvement throughout the therapy process to attentively address patients' reactions and refining the role of AI in human-AI collaboration to achieve optimal effects in the therapeutic context.

\subsection{Implications for Human-AI collaboration beyond art therapy}
\label{subsec:discuss_three}

The promising results of the HITL VA RecSys in the context of art therapy opens the door to broader applications of human-AI collaboration across various areas of healthcare. The principles demonstrated here augmenting human expertise, maintaining a balance between automation and human oversight, and prioritizing patient-centered care are applicable to numerous other therapeutic interventions. For example, similar RecSys models could be employed in personalized medicine, mental health support, and rehabilitation, where AI could assist in the selection of treatments, monitoring of progress, and adjustment of therapeutic strategies in real-time.

Moreover, the insights gained from this study contribute to the ongoing discourse on the ethical design and deployment of AI in healthcare. Ensuring that AI systems are monitored to ensure that they are risk-free is crucial, especially when they are integrated into fields like therapy. The development of AI systems that respect human autonomy, preserve the therapeutic alliance, and support the nuanced understanding that clinicians bring to their practice is essential for the responsible advancement of AI in healthcare.

\subsection{Limitations and future work}
\label{subsec:discuss_four}

While our research demonstrates the potential of Human-AI collaboration in enhancing art therapy for PICS patients, 
we also acknowledge the limitations of our current approach.
Perhaps the main limitation of our work is the dependency on 
the quality and diversity of the data we considered for evaluation. 
Although we leveraged a collection of high-quality art paintings, 
the effectiveness of AI recommendations is inherently tied to the richness of this data. 
In real-world applications, the diversity of patients’ backgrounds 
and their subjective interpretations of art 
pose a challenge to creating VA RecSys that can generalize potentially to any collection.
Future work should consider other artwork datasets,
and exploring methods to adapt AI models to individual patient profiles dynamically.

Additionally, while our study highlights the positive impact of Human-AI collaboration, particularly in terms of time efficiency and recommendation appropriateness, we also recognize limitations in the current recommendations, which occasionally may include negative paintings. On the one hand, this underscores the importance of involving a human expert in the process, as demonstrated in our study. On the other hand, it highlights the need for improving RecSys algorithms to better filter out negative paintings and reduce noise, a factor an expert noted as affecting the overall efficiency of the HITL approaches. A continuous feedback loop, for example, could be a valuable approach to optimize VA RecSys engines further.

Finally, as AI continues to evolve, so too must the frameworks that govern human-AI interaction. 
Our study lays the groundwork for understanding how AI can support art therapy, 
but future work must explore new interaction paradigms 
that further enhance the synergy between human expertise and AI capabilities. 
This includes the development of more intuitive interfaces, 
the integration of multimodal inputs that combine visual, textual, and even physiological data, 
and the continuous adaptation of AI models based on real-time feedback from both patients and therapists.

\section{Conclusion}

We have explored the potential of a human-AI collaborative approach to support PICS rehabilitation through personalized art therapy with visual exposure. Our study evaluated the efficacy of the proposed HITL approach which combines visual and multimodal VA RecSys engines with expert knowledge in the delivery of art therapy. The results demonstrate that recommendations generated through a HITL approach not only compare favourably with expert-curated ones but, in some cases, even surpass them in terms of therapeutic relevance and impact. Furthermore, the results emphasize the added value of VA  RecSys in the HITL approach, supporting therapists to save significant time and streamline the guided therapy creation process. Our findings suggest a promising future where HITL  plays a crucial role in improving personalised intervention through art therapy not only for PICS rehabilitation but also for other areas where patients need emotional support. Thus, opening an interesting avenue for a human-AI collaboration in the space of art therapy and beyond which has not yet been sufficiently explored. 

\begin{acks}
Research supported by the Horizon 2020 FET program of the European Union through the ERA-NET Cofund funding (grant CHIST-ERA-20-BCI-001)
and the European Innovation Council Pathfinder program (SYMBIOTIK project, grant 101071147).
\end{acks}

\bibliographystyle{ACM-Reference-Format}
\bibliography{refs}

\end{document}